\documentclass[final,5p,times,twocolumn]{elsarticle}
\usepackage{amssymb}
\usepackage{color}
\usepackage{amsmath}
\usepackage{multirow}
\usepackage{booktabs}
\usepackage{braket}

\usepackage{graphicx}
\usepackage[pdfstartview=FitH, colorlinks,
 linkcolor={red!60!black!},
 anchorcolor={green!80!black!},
 urlcolor={blue!80!black!},
 citecolor={blue!50!black!}]{hyperref}
\usepackage[table]{xcolor}
\journal{Science Bulletin}

\begin{document}
\begin{frontmatter}

\title{Can a pure neutron state of matter exist?}

\author{Furong Xu}
\address{State Key Laboratory of Nuclear Physics and Technology, School of Physics, Peking University, Beijing 100871, China}

\cortext[cor1]{frxu@pku.edu.cn}

 


\end{frontmatter}

Atomic nucleus is a self-organized system composed of protons and neutrons through the short-range attractive nuclear force. For protons there is a long-range Coulomb repulsion, while no Coulomb interaction exists between neutrons. Then, a naive question arises: can a pure neutron system exist? A pure neutron system seems to be more bound because there is no Coulomb repulsion between neutrons. However, no bound pure neutron system has been found except in neutron stars where neutrons are squeezed together by the huge gravitational force. 

From a simple phenomenological macroscopic point of view, an atomic nucleus can be considered as a droplet with volume, surface and Coulomb energies. Indeed, the Coulomb repulsion reduces the binding energy of the nucleus, which means that the Coulomb energy is not conducive to the binding of the nucleus. However, there exists a quantum term, called the symmetry energy proposed by Wigner, which arises from the Pauli exclusion of identical particles. In the symmetry energy term, the asymmetry in the number of neutrons and protons results in a decrease in the binding energy of the nucleus, which means that the symmetry energy prefers similar numbers of neutrons and protons to very different numbers. Therefore, according to the phenomenological liquid drop model, the Coulomb energy and symmetry energy compete to balance the numbers of neutrons and protons to make the nucleus lower in energy. 

In a microscopic framework of the shell model, neutrons and protons occupy their respective single-particle levels, and prefer their Fermi surfaces close to each other making the A-body nucleon system lower in energy. A nucleus can balance its neutron and proton numbers through the $\beta$ transition in which a neutron (proton) is converted into a proton (neutron) by the emission of an electron (positron) accompanied by an antineutrino (neutrino). The neutron and proton can be approximately treated as the same type of Fermions by introducing an isospin quantum number, e.g., defining the isospin projection $t_z=1/2$ for neutron and $-1/2$ for proton. Then, the nucleus appears with an isospin quantum number as well, similarly to the spin number of its angular momentum, with an isospin projection $T_z=\frac{1}{2}(N-Z)$ measuring the difference in the number of neutrons and protons. 

Figure 1 plots the ground-state energies of $A=4$ nuclear systems. We see that the $A=4$ energy is minimized at $T_z=0$, i.e., at the $^4$He nucleus in which the neutron number is equal to the proton number. While the $T_z=0$ $^4$He is well bound, the $T_z=-1$ $^4$Li and $T_z=+1$ $^4$H are unbound resonances which are above the $^3$He+p and $^3$H+n breakup thresholds, respectively. Therefore, a pure neutron system is not favored, though the Coulomb interaction between protons is repulsive. The many-body correlations of a nuclear system favor a certain balance in the number of neutrons and protons. 

Nevertheless, it is still possible that pure neutron systems might exist in the form of resonances. The first pure neutron system that comes to mind may be the dineutron with two neutrons (2n), which is unbound by only about 100 keV. However, the two neutrons stay in the $l=0$ s-wave with zero centrifugal barrier, and therefore no 2n resonance is possible. However, a pure four-neutron (4n) system, called the tetraneutron, was speculated about 60 years ago, and has been sought ever since. Unfortunately, no firm evidence for a bound 4n state has been obtained. Whether can the 4n system exist in the form of resonance? Indeed, an experiment performed in 2016 at RIKEN, using the double-charge-exchange (DCX) reaction $^4$He($^8$He, $^8$Be) \cite{2_PhysRevLett.116.052501}, indicates the possibility of the existence of the tetraneutron resonance, though the conclusion cannot be definitely made due to the large experimental uncertainty. Many theoretical attempts have also been made to investigate possible multineutron systems. Most theoretical investigations do not support bound multineutrons \cite{3_N.K.Timofeyuk_2003, 4_C.A.Bertulani_2003, 5_PhysRevLett.90.252501}, instead suggesting multineutron resonances \cite{6_PhysRevLett.117.182502, 7_AIP.Conf.Proc.2038.020038.(2018), 8_PhysRevLett.118.232501, 9_PhysRevLett.119.032501, 10_PhysRevC.100.054313}. Among the theories, first-principles calculations based on realistic nuclear forces are fascinating. However, one also needs to well treat the neutron correlations including the coupling to the continuum \cite{10_PhysRevC.100.054313}. 
\begin{figure}[ht]
    \includegraphics[width=0.48\textwidth]{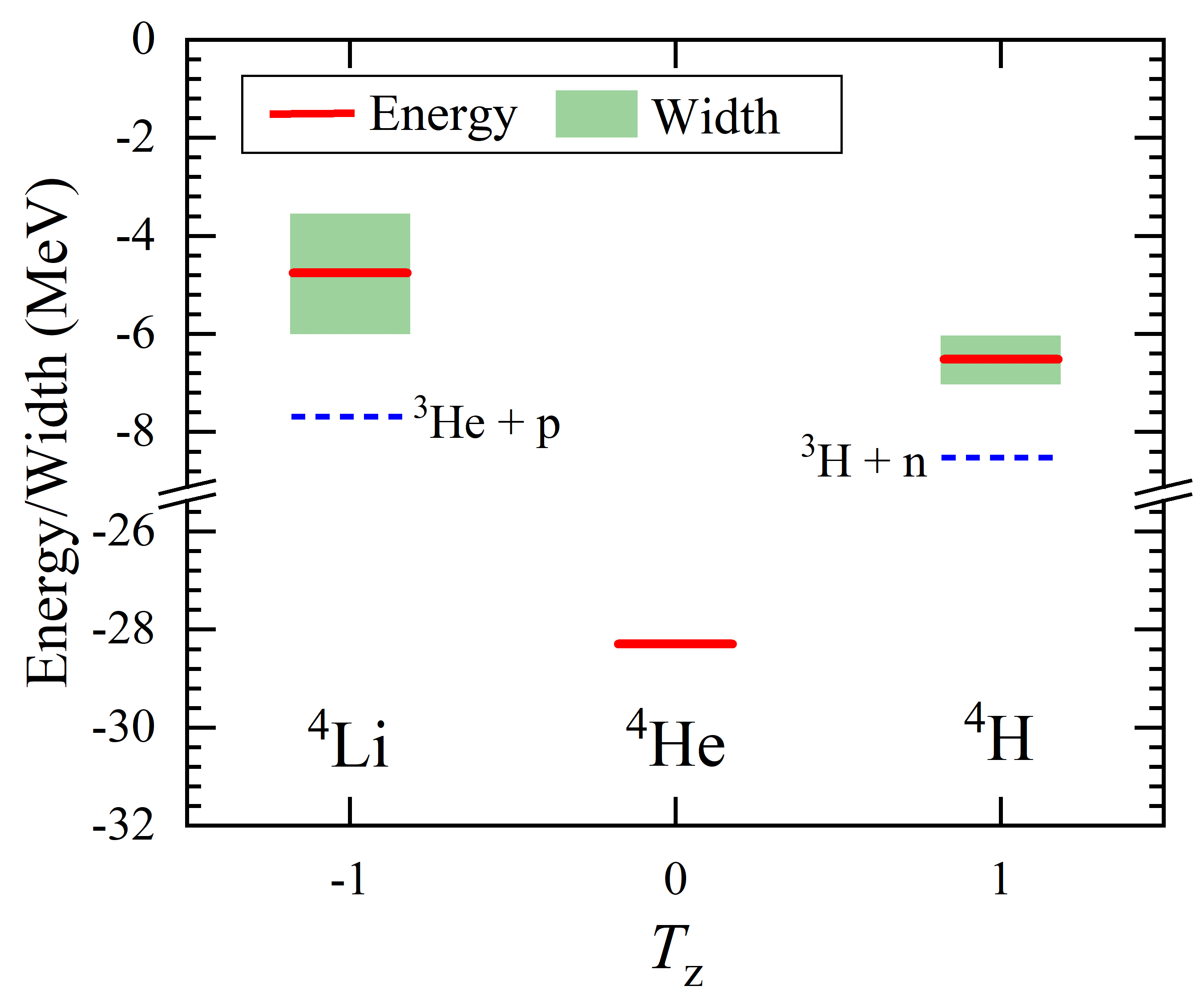}
    \caption{\label{fig1} The ground-state energies of $A=4$ nuclear systems as a function of the isospin projection $T_z$. The calculated energies and resonance widths from \cite{1_J.G.Li_PhysRevC.104.024319(2021)} are adopted because available experimental data for $^4$Li and $^4$H can be significantly different from one experiment to another (see Ref. \cite{1_J.G.Li_PhysRevC.104.024319(2021)}). The blue dashed line indicates the breakup threshold.}
\end{figure}

After the DCX reaction $^4$He($^8$He, $^8$Be) experiment done in 2016 \cite{2_PhysRevLett.116.052501}, an international collaboration team performed another fascinating experiment using the knockout reaction $^8$He(p, p$^4$He), published in Nature recently \cite{11_nature606.678(2022)}. In the experiment, the high-energy $^8$He secondary beam was projected onto a target of protons. The $^8$He nucleus has a structure of four valence neutrons outside the $\alpha$ core (i.e., $^4$He). In the quasi-elastic knockout reaction, the $\alpha$ core was removed from $^8$He by a target proton, leaving a 4n system moving on. The experiment \cite{11_nature606.678(2022)} did not directly measure the 4n system, but detected the knocked-out $\alpha$-particle and the scattered proton. By the missing mass reconstruction, the energy spectrum of the 4n system can be obtained, which gives a well pronounced peak at an energy of 2.37$\pm$0.38(statistical error)$\pm$0.44(systematic error) MeV near the 4n threshold, suggesting a resonance-like structure of the 4n system \cite{11_nature606.678(2022)}. This suggestion of a 4n resonance is in agreement with the DCX reaction 4He(8He, 8Be) experiment \cite{2_PhysRevLett.116.052501}, but with much higher experimental precisions. A resonance width was extracted, $\Gamma$=$1.75\pm 0.22$(stat.)$\pm$0.30(sys.) MeV, which corresponds to a lifetime of only $(3.8\pm 0.8)\times 10^{-22}$ seconds \cite{11_nature606.678(2022)}. 

Certainly, more experimental and theoretical studies are necessary to further understand the nature of the observed peak in the missing mass spectrum \cite{11_nature606.678(2022)} and also to exclude other possibilities, e.g., the peak manifesting itself as a possible coupling among the 4n and other ingredients surrounding the reaction channel, as commented in Refs. \cite{12_nature606.656(2022),13_arXiv.2207.07575}. A direct detection of the 4n system should be of most value. The measurement of the momentum correlation of the four neutrons can give more insights into the structure of the 4n system. Choosing different reaction mechanisms would provide additional information about the properties of the 4n system. Nevertheless, combined with the recent state-of-the-art first-principles calculations [6-10], a tetraneutron resonance is highly likely. Figure 2 displays the {\it ab initio} predictions based on no-core shell model (NCSM) \cite{6_PhysRevLett.117.182502,7_AIP.Conf.Proc.2038.020038.(2018)} and no-core Gamow shell model (NCGSM) \cite{10_PhysRevC.100.054313}, along with the energies and widths of the 4n peaks obtained in the experiments \cite{2_PhysRevLett.116.052501, 11_nature606.678(2022)}. We see that the newest experimental result published in Nature \cite{11_nature606.678(2022)} is in excellent agreement with the theoretical prediction made by the elaborated {\it ab initio} NCGSM \cite{10_PhysRevC.100.054313}. Starting from a fundamental chiral nuclear force, the NCGSM \cite{10_PhysRevC.100.054313} treated well the coupling to the continuum by using the complex-momentum Berggren ensemble. The continuum effect is important in the calculation for a resonant state. The use of natural orbitals includes more correlations \cite{10_PhysRevC.100.054313}. 
\begin{figure}[ht]
    \includegraphics[width=0.48\textwidth]{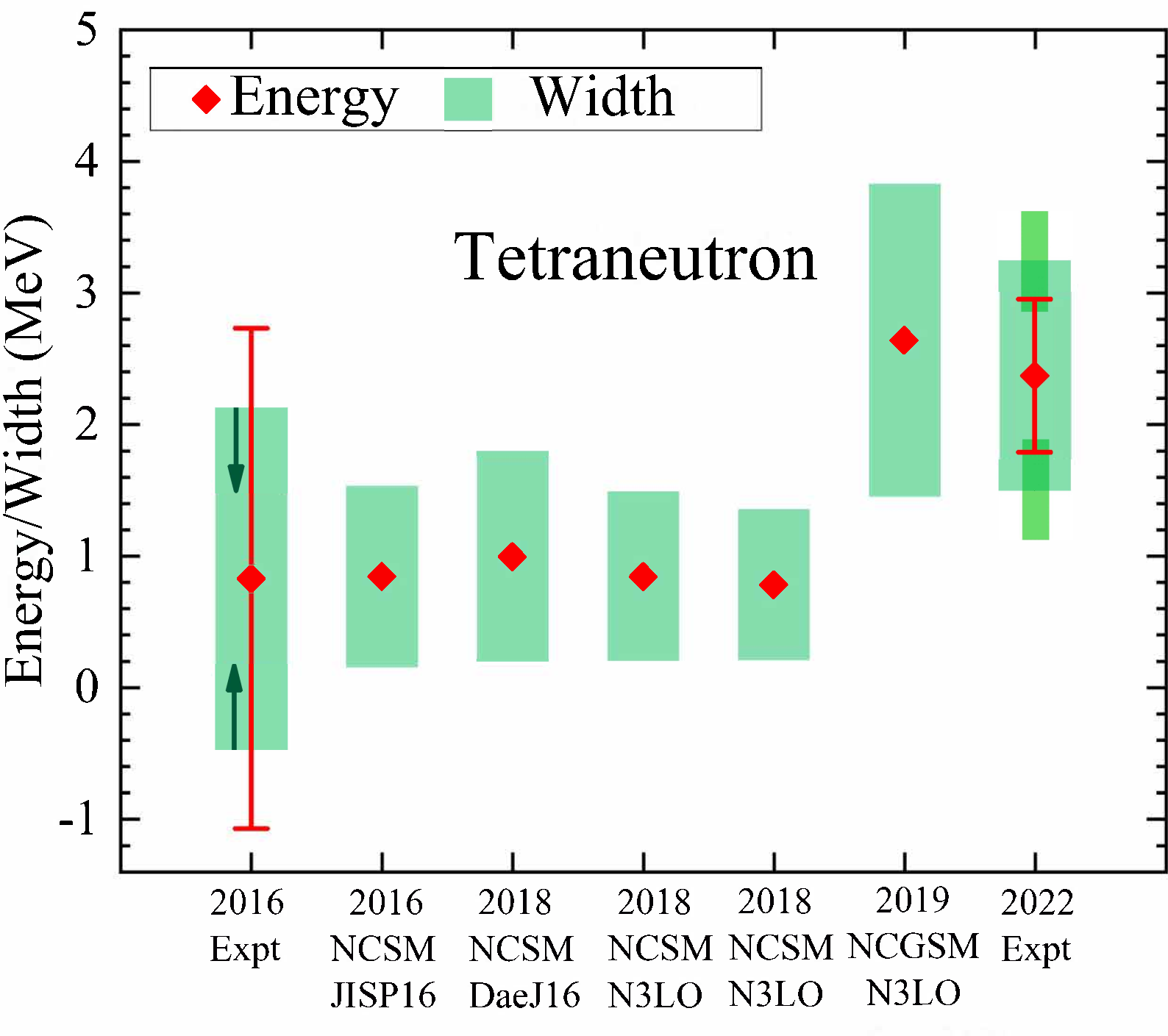}
    \caption{\label{fig2} Experimental \cite{2_PhysRevLett.116.052501, 11_nature606.678(2022)} and predicted \cite{6_PhysRevLett.117.182502, 7_AIP.Conf.Proc.2038.020038.(2018), 10_PhysRevC.100.054313} resonance energy and width of the 4n system, with the horizontal axis indicating the year when the experimental or calculated result was published. In the year-2022 experimental data \cite{11_nature606.678(2022)} the narrower green shadow indicates the experimental error in the width \cite{11_nature606.678(2022)}, while the year-2016 experiment \cite{2_PhysRevLett.116.052501} only estimated an upper limit on the width, indicated by blue arrows. JISP16, DaeJ16 and N3LO indicate the realistic nuclear forces used in NCSM \cite{6_PhysRevLett.117.182502, 7_AIP.Conf.Proc.2038.020038.(2018)} and NCGSM calculations \cite{10_PhysRevC.100.054313}.}
\end{figure}

The {\it ab initio} calculations \cite{8_PhysRevLett.118.232501, 10_PhysRevC.100.054313} have even showed that a three-neutron (3n) resonance is possible with a lower energy and narrower width than the 4n resonance. Though it has been commented that the 3n resonance may be less likely to exist due to the odd number of nucleons and thus weaker binding \cite{11_nature606.678(2022)}, one probably should not consider it with the pairing mechanism of even-number nucleon systems. One might treat it as a whole with full three-body correlations and three-nucleon force. For the 4n system, a realistic four-nucleon force may also need to be considered.

Multineutron resonances provide a unique playground for the studies of neutron-neutron interaction and many-body correlations of pure neutron systems. One can study neutron-proton and proton-proton interactions by neutron-proton and proton-proton scatterings, respectively, but neutron-neutron scattering is not an idea to study the neutron-neutron interaction because no pure neutron target is available. Further studies of the pure neutron states of matter are necessary to answer the fundamental questions of nuclear physics and the impacts, e.g., on the nuclear processes of nuclear astrophysics.

\bigskip



\vspace{0.3cm}
{\noindent \bf Acknowledgements}
\vspace{0.3cm}

Dr. Jianguo Li at Institute of Modern Physics, Chinese Academy of Sciences, is acknowledged for preparing the figures and useful discussions. The author thanks the support from the National Natural Science Foundation of China (11835001, 11921006, and
12035001).

  \bibliographystyle{elsarticle-num_noURL}
\bibliography{reference}


\end{document}